\documentclass[11pt]{article}
\begin{document}
	\def\ba{\begin{eqnarray}}
	\def\ea{\end{eqnarray}}
	\def\w{\wedge}
	\def\d{\mbox{d}}
	\def\D{\mbox{D}}

\begin{titlepage}	
\title{Dark range of $\omega$ in Brans-Dicke gravity}
\author{Tekin Dereli\footnote{tdereli@ku.edu.tr},     
Yorgo \c{S}eniko\u{g}lu\footnote{ysenikoglu@ku.edu.tr}
\\
{\small  Department of Physics, Ko\c{c} University, 34450 Sar{\i}yer-\.{I}stanbul, Turkey}  }

\vskip 1cm 

\date{ }

\maketitle	
	
\vskip 2cm 

\begin{abstract}
\noindent The variational field equations of Brans-Dicke scalar-tensor theory of gravitation are presented in a Riemannian and non-Riemannian setting in the language of exterior differential forms over 4-dimensional spacetime. In Rosen coordinates, the equations of motion of non-spinning test masses are considered in a gravitational plane wave setup and detailed to interpret a scalar field $\alpha$ as a candidate for dark energy for negative values of the Brans-Dicke parameter  $-3/2 < \omega < 0$ .

\vskip 2cm

\noindent PACS numbers: 04.50.Kd, 04.50.-h
\end{abstract}
\end{titlepage}

\newpage

 \section{Introduction}
General relativity is a geometrical theory of gravitation that is determined by the metric of spacetime. Brans-Dicke theory of gravity\cite{B1, B2, B3}, a scalar-tensor theory, introduces an adjustment to the initial framework of Einstein in order to include Mach's Principle by including a scalar field $\phi$. The astrophysical observations supplied to scientist with the construct of low energy string theories with scalar fields are very efficient in depicting the large scale structure of spacetime and subatomic physics; so it is very reasonable to recognize them also at large scales. 

Brans-Dicke theory covers the geodesic postulate; the motion of a test mass is determined by the geodesics with the gravitational effects of all particles embedded in the metric and the associated Levi-Civita connection. 
The assumption that the matter lagrangian is independent of the scalar field $\phi$ secures the equivalence principle.
The string unification of gravity and all quantum powers make researchers think on the property and reasonableness of the theory of gravitational interaction being depicted, by geometry alone.
There are several hints that a non-Riemannian setting of spacetime may contribute a refined and coherent way to describe gravitational fields\cite{B4,B5,B6,B6-a}. A first order variation of the Brans-Dicke Lagrangian implies that the spacetime connection can admit a non-vanishing torsion. The field equations obtained are equivalent to up to a shift in the coupling constant \cite{B7}. 

Under the influence of gravity alone, non-spinning test masses follow geodesic equations of motion in a Riemannian spacetime and autoparallels of a connection with torsion \cite{B8} in a non-Riemannian geometry. The existence of other matter fields will reshape the geometry of spacetime, and in the presence spinorial matter, torsion will be picked up by differential equations \cite{B9,B10}. 
In this article, the gradient of the scalar field determines the space-time torsion, and a particular choice of the scalar field applied to parallely propagating gravitational plane waves will reduce the field equations to the vacuum Einstein equations.  In particular, we are motivated on the interpretation in a Riemannian and non-Riemannian geometry. Explicitly, the Brans-Dicke parameter and the autoparallels of the connection with torsion which differ from the geodesic equations of motion will help us understand the coherence of the explanation.

The article is organized as follows: in Section 2, we present the Brans-Dicke theory of gravity with the gravitational field equations and the equations of motion of a non-spinning test mass in a non-Riemannian setting in the language of exterior differential forms. Section 3 details the plane wave solutions in Rosen coordinates and a particular choice of scalar field that reduce the field equations to source-free Einstein equations. Section 4 is devoted to concluding remarks. 
\bigskip

\section{Brans-Dicke Theory Field Equations}

\noindent 
Let  $M$ be a 4-dimensional spacetime manifold, the field equations are determined from the action $I=\int_{M} \mathcal{L},$ where the Brans-Dicke Lagrange density 4-form

\begin{equation}
{\mathcal{L}} = \frac{\alpha^2}{2}R_{ab} \wedge *(e^a \wedge e^b) -\frac{c}{2} d\alpha \wedge *d\alpha.
\end{equation}

\noindent $\{e^a\}$ 's denote the co-frame 1-forms in terms of which the space-time metric is given by $g=\eta_{ab}e^{a} \otimes e^b$ with $\eta_{ab}=diag(-+++)$. ${}^*$ stands for the Hodge map. The volume form is $*1 = e^0 \wedge e^1 \wedge e^2 \wedge e^3$. We write the Brans-Dicke scalar field as $\phi=\alpha^2$ for convenience. $\{{\omega}^{a}_{\;\;b}\}$ are the connection 1-forms that satisfy the Cartan structure equations
  \begin{equation}
  de^a + \omega^{a}_{\;\;b}  \wedge e^b = T^a
  \end{equation}
  with the torsion 2-forms $T^a$ and  
  \begin{equation}
 d \omega^{a}_{\;\;b} + \omega^{a}_{\;\;c}  \wedge \omega^{c}_{\;\;b} =R^{a}_{\;\;b}   
  \end{equation}
with the curvature 2-forms $R^{a}_{\;\;b}$ of spacetime. 

We vary the action with respect to the co-frame 1-forms $e^a$, connection 1-forms $\omega^{a}_{\;\;b}$ and
the scalar field $\alpha$. The corresponding coupled field equations turn out to be $(c \neq 0)$ in the non-Riemannian description:
\begin{eqnarray}
-\frac{\alpha^2}{2} R^{bc} \wedge *(e_a \wedge e_b \wedge e_c) &=& c\hspace{1mm}\tau_a[\alpha], \nonumber \\ 
T^a &=& e^a \wedge \frac{d \alpha}{\alpha} ,   \nonumber \\ 
c d*d\alpha^2 &=& 0;  \label{eq:FE-NR}
\end{eqnarray}
where
\ba
\tau_a[\alpha]=\frac{1}{2} \left (  \iota_a d\alpha *d\alpha + d\alpha\wedge \iota_a *d\alpha  \right ) \equiv T_{ab} *e^b
\ea
are the energy-momentum 3-forms of the scalar field and $\iota_a$ denote the interior products that satisfy 
$\iota_a e^b = \delta^{b}_{\;\; a}$.

Denoting the Levi-Civita connection 1-forms 
$\{{\hat{\omega}}^{a}_{\;\;b}\}$, fixed in a unique way by the metric tensor through the Cartan structure equations
\begin{equation}
de^a + {\hat{\omega}}^{a}_{\;\;b} \wedge e^b = 0,
\end{equation}
we obtain the equivalent field equations to the above field equations (\ref{eq:FE-NR}) in the Riemannian description as:
\begin{eqnarray}
-\frac{\alpha^2}{2}{\hat{R}}^{bc} \wedge *(e_a \wedge e_b \wedge e_c) &=& \frac{(c-6)}{2} \left (  \iota_a d\alpha *d\alpha + d\alpha \wedge \iota_a *d\alpha  \right ) \nonumber \\
& & +\hat{D}(\iota_a*d\alpha^2),  \nonumber \\
c d*d\alpha^2 &=&0,
\end{eqnarray} 
where $\hat{D}$ denotes the covariant exterior derivative with respect to the Levi-Civita connection.

Then the coupling constant $c$ can be identified in the classical Brans-Dicke theory by 
\ba
c=4\omega +6
\ea
where $\omega$ denotes the conventional Brans-Dicke coupling constant.
\section{Plane Wave Solutions}
\noindent 
We consider the plane fronted gravitational wave metric in Rosen  coordinates $(u,v,x,y)$  to solve the Brans-Dicke field equations\cite{B11,B12}:
\ba
g = 2\d u \otimes \d v +  \frac{\d x \otimes \d x}{f(u)^2} + \frac{\d y \otimes \d y}{h(u)^2}.   \label{eq:PP-metrik0}
\ea     
We further take a scalar field 
\ba
\alpha&=& \alpha(u).
\ea
\noindent
We note that the scalar field equation to the system is again identically satisfied.
Working out the expressions for the curvature, torsion and  the  scalar field stress-energy-momentum tensors, the Einstein field equations to be solved 
reduce to the following second order differential equation: 
\ba
\frac{f_{uu}}{f} -\frac{2f_u^2}{f^2} + \frac{h_{uu}}{h} -\frac{2h_u^2}{h} -\frac{2\alpha_{uu}}{\alpha}+\frac{4\alpha_u^2}{\alpha^2}=\frac{c\alpha_u^2}{\alpha^2}. \label{eq:FieldEq}
\ea
\noindent
For a particular choice of $\alpha$ such that
\ba
\alpha(u)= \Big((\omega+1)(C_1u+C_2)\Big)^{\frac{1}{2(\omega+1)}}, \label{eq:spe-alpha}
\ea
where $C_1$ and $C_2$ are arbitrary integration constants, all the $\alpha$-dependent terms in (\ref{eq:FieldEq}) drop out and we recover the source-free Einstein field equation:
\ba
\Big(\frac{f_{uu}}{f} -\frac{2f_u^2}{f^2} + \frac{h_{uu}}{h} -\frac{2h_u^2}{h}\Big)=0.
\ea
If we study the articles of Brans and Dicke, we notice that the right hand side of the field equations in a Riemannian setting is composed of two terms\cite{B11-a}: one term contains the Brans-Dicke parameter multiplying the scalar field stress-energy tensor; that comes from the kinetic term of the scalar field in the action, while the other does not contain any parameter. In fact it comes from the geometry of the spacetime itself, and it cannot be seen explicitly in the non-Riemannian description. We emphasize that there exists such parameter values where these two terms cancel each other out and the field equations reduce to the source-free Einstein equations as we have shown above. Thus in certain spacetime geometries the geodesic equations of motion are not affected by the scalar field at all in the conventional Riemannian approach. There is a scalar field but 
non-spinning test masses are not influenced by it. We have a conceptual problem in the conventional Riemannian interpretation whereas in the non-Riemannian description there is none.

Let us decorate this observation by explicitly giving 
\noindent
the geodesic equations as functions of proper time $\tau$ and constants of motion $p_u, p_x, p_y$ neatly as \cite{B12} 
\ba
u(\tau)=u(0)+\frac{p_u\tau}{m} , \nonumber \\
v(\tau)=v(0)-\frac{m\tau}{2p_u}-\frac{p_x^2}{2p_u^2}\int_0^\tau f(u)^2du-\frac{p_y^2}{2p_u^2}\int_0^\tau h(u)^2du ,\nonumber\\
x(\tau)=x(0)+\frac{p_x}{p_u}\int_0^\tau f(u)^2du, \quad 
y(\tau)=y(0)+\frac{p_y}{p_u}\int_0^\tau h(u)^2du.
\ea
Similarly the autoparallel equations are given as 

\ba
u(\tau)=u(0)+\int_{0}^{\tau}\frac{d\tau}{\alpha}, \nonumber\\ 
v(\tau)=v(0)-\frac{1}{2}\int_{0}^{\tau}\alpha(u)^2du-\frac{p_x^2}{2m^2}\int_{0}^{\tau}f^2(u)du-\frac{p_y^2}{2m^2}\int_{0}^{\tau}h^2(u)du, \nonumber\\ 
x(\tau)=x(0)+\frac{p_x}{m}\int_{0}^{\tau}f^2(u)du, \quad
y(\tau)=y(0)+\frac{p_y}{m}\int_{0}^{\tau}h^2(u)du,
\ea
where in the former case a test mass does not interact with the field and in the latter it clearly does.

\vspace{2mm}
\noindent
Taking the Brans-Dicke field equations with torsion and looking at solutions, we have found a particular gravitational plane wave solution. The metric is the source-free (vacuum) Einstein metric and the terms coming from torsion and the stress-energy momentum tensor cancel out and the scalar field equation is satisfied. In the non-Riemannian description, we do not have any issues, on the right hand side, the energy momentum tensor is different than zero and positive definite. The left hand side of the field equations, the geometry is affected by torsion, given by the gradient of the scalar field. When we look at the autoparallel equations they are not the source-free Einstein geodesic equations. We find different equations of motion as expected. Had we insisted on the classical Brans-Dicke interpretation, this solution would be a solution again. But what kind of a solution? The left hand side ensures the Einstein vacuum equations while at the right hand side, there would have been solutions such that the improvement term and the energy momentum tensor cancel out and we would have interpreted them as "ghosts". The fundamental problem is when we consider the geodesic hypothesis. The presence of the scalar field is assumed not to affect the geodesics in this case. We have a ghost field in a sense analogous to the ghost neutrino problem\cite{B13,B14},  The theory is not scale invariant, the metric is the vacuum Einstein metric and the equations of motion are the geodesics that are implied by the Einstein field equations, and the test masses do not see the scalar field. In the standart interpretation, this is an issue but in an interpretation with torsion, it is not. This we believe is a strong case for the non-Riemannian interpretation of Brans-Dicke gravity.
\section{Concluding Remarks}
In this article, we studied a gravitational plane wave solution of the Brans-Dicke field equations to point the consistentency interpretation of the non-Riemannian setting. The spacetime geometry is relaxed to admit torsion, depending on the gradient of the scalar field. A particular choice of the scalar field $\alpha$ reduces the field equations to the source-free Einstein field equations. The clarification on the sign is of utmost importance. There are two terms that cancel out in the right hand side, one is the shifted energy-momentum tensor and another term that comes from the geometry. What we explicitly want to show is the existence of such a configuration for $0 < c < 6$ i.e. for a negative $\omega$, specifically $-\frac{3}{2}<\omega<0$. This may well be a dark energy candidate. On the other hand, we have detailed that if we try to explain the theory in a Riemannian description, we come to a pathological case where there is a scalar field present that cancels, in a way, the right hand side of the Einstein equations and that in the same geodesic hypothesis the scalar test mass does not interact with this scalar field. If we adopt a more consistent hypothesis, that scalar test masses should follow autoparallels of the connection with torsion, then the identification of the energy-momentum tensor and the equations of motion are coherent. Torsion explicitly affects as it should, the equations of motion of the scalar test masses. Brans-Dicke field equations and the equivalence principle should be given in a non-Riemannian setting to be viable, modest and coherent. That is why we have discussed this in a particular plane wave example with a negative Brans-Dicke parameter $\omega$, dark $\omega$, which can depict some form of dark energy.

\section{Acknowledgement}
Y.\c{S}. is grateful to Ko\c{c} University for its hospitality and partial support.

\vskip 1cm

{\small 

\bibliography{references}

\begin{thebibliography}{99}

\bibitem{B1} C.H.Brans,R.H.Dicke,{\sl  Mach's principle and a relativistic theory of gravitation},Phys.Rev.{\bf 124},925(1961)   

\bibitem{B2} R.H.Dicke,{\sl Mach's principle and invariance under transformation of units},Phys.Rev.{\bf 125},2163 (1962) 

\bibitem{B3} C.H. Brans,{\sl Mach's principle and a relativistic theory of gravitation II},Phys.Rev.{\bf 125},2194 (1962)

\bibitem{B4} R.W.Tucker,C.Wang,{\sl Black holes with Weyl charge and non-Riemannian waves},Class.Q.Grav.{\bf 12}(1995)2587 

\bibitem{B5} P.Teyssandier,R.W.Tucker,{\sl Gravity,gauges and clocks},Class.Q.Grav.{\bf 13}(1996)145

\bibitem{B6}  P.Teyssandier,R.W.Tucker,C.Wang,{\sl On an interpretation of non-Riemannian gravitation},Acta Phys.Polon.{\bf B29}(1998)987

\bibitem{B6-a} T.Dereli, Y.Senikoglu, {\sl Non-Riemannian description of Robinson-Trautman spacetimes in Brans-Dicke gravity}, Int. J. Mod. Phys. {\bf D 28}, 1950070 (2019)
\bibitem{B7}  T.Dereli,R.W.Tucker,{\sl Weyl scalings and spinor matter interactions in scalar-tensor theories of gravitation},Phys.Lett.{\bf B110}(1982)206

\bibitem{B8}T.Dereli,R.W.Tucker,{\sl On the detection of scalar field induced space-time torsion},Mod.Phys.Lett.{\bf A17}(2002)421

\bibitem{B9} D.Burton,T.Dereli,R.W.Tucker,{\sl On the motion of matter in gravitational fields}, in {\bf Symmetries in Gravity and Field Theory}, V.Aldaya, J.M.Cer-vero,Y Pilar Garcia (Editors) ( Ediciones Universidad Salamanca, 2004) pp.237-249  [arXiv:gr-qc/0107017]

\bibitem{B10} D.A.Burton,R.W.Tucker,C.H.Wang,{\sl Spinning particles in scalar-tensor gravity},Phys.Lett.{\bf A372}(2008)3141

\bibitem{B11} J.B.Griffiths,J.Podolsky,{\bf Exact Space-Times In Einstein's General Relativity}(Cambridge U.P.,2009),Chapter:17

\bibitem{B12} T.Dereli,Y.Senikoglu, {\sl Gravitational Plane Waves in a Non-Riemannian Description of Brans-Dicke Gravity}, [arXiv:gr-qc/1904.11219]

\bibitem{B11-a} C.G.Callan Jr.,S.Coleman,R.Jackiw {\sl A new improved energy momentum tensor},Ann.Phys. {\bf 59}(1970)42-73 

\bibitem{B13} T.M.Davis,J.R.Ray {\sl Ghost neutrinos in general relativity}, Phys.Rev.{\bf D9}(1974)334

\bibitem{B14} T.Dereli,R.W.Tucker {\sl Exact neutrino solutions in the presence of torsion},Phys.Lett.{\bf A 82},(1981)229
\end{thebibliography}
\bibliographystyle{iopart-num}
}
\end{document}